\documentclass{aa}
\usepackage{fixltx2e}
\usepackage{amsmath}
\usepackage{aas_macros}
\usepackage{subfigure}
\usepackage{graphicx}
\usepackage{txfonts}
\usepackage{subfigure}
\usepackage{natbib}
\usepackage{siunitx}
\DeclareSIUnit\parsec{pc}
\DeclareSIUnit\gauss{G}

\newcommand{\ipt}{\omega_{pe}^{-1}}
\newcommand{\epf}{\omega_{pe}}
\begin{document}
   \title{Energy loss of intergalactic pair beams: Particle-in-Cell simulation}
   \titlerunning{Energy loss of intergalactic pair beams}

   \author{A. Kempf
          \inst{1}
          \and
          P. Kilian
          \inst{2}
          \and
          F.~Spanier
          \inst{2}
          }

   \offprints{A. Kempf, \\ \email{ank@tp4.rub.de}}

   \institute{Lehrstuhl für Theoretische Physik IV, Ruhr-Universität Bochum, 44780 Bochum, Germany
   \and
   Centre for Space Research, North-West University, 2520 Potchefstroom, South Africa}
   \date{Received ... , accepted ...}
  \abstract
   {}
   {The change of the distribution function of electron-positron pair beams determines whether \si{\giga\electronvolt} photons can be produced as secondary radiation from \si{\tera\electronvolt} photons. We will discuss the instabilities driven by pair beams.}
   {The system of a thermal proton-electron plasma and the electron-positron beam is collision free. We have, therefore, used the Particle-in-Cell simulation approach. It was necessary to alter the physical parameters, but the ordering of growth rates has been retained.}
   {We were able to show that plasma instabilities can be recovered in particle-in-cell simulations, but their effect on the pair distribution function is negligible for beam-background energy density ratios typically found in blazars.}
   {}

   \maketitle

\section{Introduction}

Active Galactic Nuclei (AGN) at cosmological distances may emit photons with energies well above \SI{1}{\tera\electronvolt}, but the mean free path of these photons is limited due to the interaction with the extragalactic background light (EBL).
Even though there are some uncertainties with regard to the spectrum of the EBL, the mean free path for photons with energies $E_\gamma$ above a \si{\tera\electronvolt} can be estimated to be $D \approx \SI[parse-numbers=false]{80 (E_\gamma / \SI{10}{\tera\electronvolt})^{-1}}{\mega\parsec}$ \citep{2009PhRvD..80l3012N}.
The interaction of high energy photons with the EBL produces electron-positron pairs, where each of the leptons carries approximately half of the energy of the incident gamma-ray photon.\\
These very-high-energy pair beams are subject to several physical processes: deflection by the intergalactic magnetic field \citep{2010Sci...328...73N}, initiation of plasma instabilities in the thermal background medium \citep{Broderick_2012,2012ApJ...758..101S,2012ApJ...758..102S,2013ApJ...770...54M} or inverse Compton scattering off the Cosmic Microwave Background (CMB) \citep{2004A&A...413..807K}.
These processes affect observable quantities: While the latter process should yield photons upscattered to \si{\giga\electronvolt} energies, the other processes will add to the heating of the intergalactic medium instead.\\
Observations by the Fermi satellite \citep{2009ApJ...707.1310A} have opened up the window for observations in the \si{\giga\electronvolt} photon range. This enables observers to look for a possible \si{\giga\electronvolt} signal from secondary radiation in distant AGN.
With the spectrum of the EBL being known to some degree and AGN being observed at \si{\tera\electronvolt} energies, the tentative spectrum of \si{\giga\electronvolt} photons could be calculated.
The question at hand is which process dominates the evolution of the electron-positron pair beam.
If magnetic field deflection is the dominating process, the beam will dissolve and the inverse Compton signal will be reduced substantially.
On the other hand, the evolution of a plasma instability might be the fastest process.
In that case, the question becomes whether the change of the particle distribution function is sufficient to prevent any of the other processes.

The absence of a \si{\giga\electronvolt} signal in certain blazars has been attributed solely to magnetic field deflection in order to determine the strength of the intergalactic magnetic field \citep{2010Sci...328...73N}.
However, this approach does not take into account any effects arising from plasma instabilities.
Due to the highly non-thermal distribution function of electrons in the intergalactic medium (IGM), there is a lot of potential for plasma instabilities to develop.

While the growth rates of plasma instabilities caused by ultra-relativistic electrons can be calculated at the onset of the instability, later non-linear phases require several assumptions and approximations in an analytic treatment.
The ultimate outcome is determined, in particular, by the interaction of turbulence with the initial distribution function and other instabilities.
Unfortunately, simulations cannot reflect the extreme parameter space of physical reality.
Therefore, a simplified system needs to be constructed carefully in order to capture all the essential physical processes of the phenomena to be examined.
In previous Particle-in-Cell simulations \citet{Sironi_2014} studied a wide range of density ratios $\alpha$ and beam Lorentz factor $\gamma_b$ to allow for a reliable extrapolation.
In this article we argue that the behavior of the system is governed principally by the energy density of the beam compared to the background energy density.
Consequently, quantities like the high beam gamma factor and high density ratios need not be as problematic for PiC simulations as long as the full range of $\epsilon$ is studied, covering values both larger and smaller than unity. The latter case is especially interesting, because it is not realized for any of the parameters discussed by \citet{Sironi_2014}.

\section{Physical parameters}

The physical setup of the system consists of hot, low-density ionized gas in thermal equilibrium and a high-intensity photon population with energies up to 10 TeV as well as a second photon population in the IR-UV range (the EBL).

The parameters of the IGM are, to some degree, uncertain, but there is some consensus that the temperatures are in the range of \SI{e4}{\kelvin} in cosmic voids \citep{1997MNRAS.292...27H} to \SI{e7}{\kelvin} in the intra-cluster medium \citep{2015A&A...575A..37M}. Following the approach by \citet{2012ApJ...752...23C} we have adopted a value of \SI{e6}{\kelvin} and a density of $n_\text{background} = \SI{e-7}{\per\cubic\centi\meter}$, which seems to be a good approximation when AGN heating plays a role. Since we are dealing with the interaction of pair beams emanated from AGN, it is a sensible approach to also take into account the heating effects.
The magnetic field is hard to determine because it is too weak to be observable through effects such as Zeeman splitting or Faraday rotation that are use to measure the strength stronger, solar magnetic fields.
This leaves the wide range between primordial magnetic fields that have not been amplified by dynamo processes and are as weak as $\SI{e-24}{\gauss}$ and an upper limit of about $\SI{e-9}{\gauss}$ \citep{Kronberg_1994} based on quasar observations.
More recently the spectrum \citep{Essey_2011} and surrounding halo \citep{Chen_2015} of distance AGN have been used to narrow this down to the range $\SI{e-17}{\gauss} \dots \SI{3e-14}{\gauss}$.
This is consistant with the value of  $\SI{e-15}{\gauss}$ favoured by \citep{Ando_2010}.

From the AGN photons and the EBL photon field the distribution function of electrons in the beam can be calculated. This will typically produce a power-law of the pairs resembling the power law of the gamma-rays from the source \citep{2012ApJ...758..102S}. Since the minimal energy to produce pairs is already at TeV energies, the width of the production spectrum is typically very small.

Due to the limited sensitivity of Air Cerenkov Telescopes and the EBL absorption, it is not possible to make definite statements about the upper end of the AGN spectrum.
As soon as the EBL has attenuated the AGN spectrum enough to make it fall under the telescope's sensitivity only models may give a clue on that part of the spectrum.

In order to describe the system, we make use of the parameters $\alpha = n_\text{jet}/n_\text{background}$ and $\epsilon = e_\text{jet}/e_\text{background}$ where $e$ is the energy density ($e_\text{jet} = n_\text{jet} \gamma m_e c^2$ and $e_\text{background}=n_\text{background}k_B T$). Typical parameters for the pair beam as derived by \citet{2012ApJ...758..101S} are $\gamma = 10^6$ and $n_\text{jet}=\SI{e-22}{\per\cubic\centi\meter}$. The actual parameters can differ over several orders of magnitude due to the extremely different AGN, but also on the EBL spectrum at the redshift of the source.

\section{Methods}

The simulations presented in this paper are prepared using the particle-in-cell (PiC) \citep{HockneyEastwood_1988} code ACRONYM \citep{Kilian_2011}.
A simple two-dimensional Cartesian grid topology (2D3V) with periodic boundary conditions in both spatial directions is used.
Moreover, the electric and magnetic field information is stored according to the standard Yee \citep{Yee_1966} scheme.
In addition, the current density is calculated following the Esirkepov scheme \citep{Esirkepov_2001}.
Current and charge density are deposited on the grid via a triangular shaped cloud (TSC) form factor, providing a good balance of computational speed and numerical accuracy \citep{Kilian_2013}.
In the electromagnetic case, Maxwell's equations are solved with an explicit second order leap-frog scheme.
For the electrostatic case, the electric field is obtained as a solution to Poisson's equation provided by a Fourier solver.
Field interpolation to particle locations proceeds with the TSC form factor used above.
The relativistic equation of motion is applied through the implicit method described by Vay \citep{Vay_2008}, although the standard Boris push \citep{Boris_1970} proved sufficient during testing.

In order to suppress unphysical field fluctuations arising from the grid Cherenkov instability, a Friedmann filter with a filtering parameter $\Theta = 0.3$ is introduced in the electromagnetic case \citep{Greenwood_2004}.
Furthermore, the current density is filtered with a spatial binomial filter including a compensation pass.

\section{Numerical setup}

\begin{table}
	\centering
	\begin{tabular}{|l | r|}
		\hline
		$T$ &
		$\SI{1.0e6}{\kelvin}$\\
		$\omega_{\mathrm{p, e}}$ &
		$\SI{17}{\radian\per\centi\meter}$\\
		$n_\mathrm{e}$ &
		$\SI{1.0e-7}{\per\cubic\centi\meter}$\\
		$\Delta t$ &
		$\SI{4.4e-3}{\second}$\\
		$\Delta x$ &
		$\SI{3.0e8}{\centi\meter}$\\
		$\Delta x / (c \omega^{-1}_{\mathrm{p, e}})$ &
		$\num{0.125}$\\
		$\gamma$ &
		$\num{10}$\\
		$\alpha = n_{\mathrm{jet}} / n_{\mathrm{bg}}$ &
		$\num{2.5e-5}$\\
		$\epsilon =  (\alpha \gamma m_\mathrm{e} c^2) /k_\mathrm{B} T $ &
		$O(\num{1})$\\
		$N_\mathrm{x} \times N_\mathrm{y}$ &
		$1024 \times 1024$\\
		particles / cell &
		$\num{100}$\\
		\hline\hline
	\end{tabular}
	\caption{Parameters of the basic simulation}
	\label{tab:sim-para}
\end{table}

All simulations are performed with the following setup.
Four particle populations are distributed homogeneously throughout the simulation volume with equal macro-particle numbers per cell.
The background consists of electrons as well as protons of natural mass ratio at thermal equilibrium at temperature $T$.
A given number density $n_\mathrm{e} = n_\mathrm{p}$ is achieved by scaling the macro-particles appropriately, as required for a PiC simulation.
The pair beam consists of positrons and electrons with  number density $\alpha n_\mathrm{e}$ each.
Its distribution is a Maxwellian of temperature $T$ with $v\text{th}=\sqrt{\frac{1}{2}\frac{k_B T}{m}}$ per dimension, drifting along the x-axis with a relativistic speed given by $\gamma_{b}$.
This particle setup effects no net charges or currents.
Initially, electric and magnetic field components are set to zero.
Moreover, the cell size $\Delta x$ is chosen small enough to resolve the electron inertial length with several cells..
Lastly, the time step size $\Delta t$ is given by the CFL criterion.
Parameters for a basic simulation run are given in table \ref{tab:sim-para}.

Below, simulations of varying beam Lorentz factor $\gamma$, density ratio $\alpha$ and energy density ratio $\epsilon$ are presented.
Also, comparison simulations with larger box sizes and different simulation sizes are provided.

In \citet{2012ApJ...752...23C} four relevant processes have been identified: Growing electrostatic fluctuations and aperiodic fluctuations which convert beam energy into kinetic energy of the background plasma and the modulational instability as well as non-linear Landau damping, which turn the plasma energy into heat.\\
Their respective growth rates are given by:\\
Electrostatic fluctuation
\begin{equation}
  \gamma_\text{E,max} = 1.6\cdot 10^{-6} N_7^{1/6} n_{22}^{1/3} \Gamma_6^{-1/3}(1-\beta_1^2\cos \Theta)^{1/3} \text{Hz}
\end{equation}
Aperiodic (Weibel) fluctuations
\begin{equation}
  \gamma_\text{W,max} = 8\cdot 10^{-10}\frac{\beta_1 n_{22}^{1/2}}{\Gamma_6^{1/2}}\text{Hz}
\end{equation}
Modulation instability
\begin{equation}
  t = \frac{\log(2.6\cdot 10^6 T_4^{1/2} N_7^{-1/2})}{2 \gamma_\text{E,max}}
\end{equation}
Non-linear Landau damping
\begin{equation}
  \tau_R \leq 8.9\cdot 10^5 \frac{\Gamma_6^{4/3}}{N_7^{1/2}T_4^{4/3}}\text{yr}
\end{equation}
The parameters $N_7$, $n_{22}$, $T_4$ and $\Gamma_6$ describe the physical system with typical numbers, where $n_\text{bg} = N_7\SI{e-7}{\per\cubic\centi\meter}$, $n_\text{jet}=n_{22}\SI{e-22}{\per\cubic\centi\meter}$, $\gamma = 10^6 \Gamma_6$ and $T=\SI{e4}{\kelvin}$. In our scenario $\alpha$ and $\epsilon$ are the relevant quantities. Here $N_7=1$ and $T_4=100$ are constants. Rewriting the above equations using our set of variables yields
\begin{align}
  \Gamma_6            &= \frac{\epsilon}{\alpha} \frac{k_B 10^4 T_4}{10^6 m_e c^2} = 1.69 10^{-10} \frac{\epsilon}{\alpha}\\
  n_{22}              &= \frac{10^{-7} N_7}{10^{-22} n_{22}} \alpha = 10^{15}\alpha\\
  \gamma_\text{E,max} &= 289.6 \alpha^{2/3} \epsilon^{1/3}(1-\beta_1^2\cos \Theta)^{1/3} \text{Hz}\\
  \gamma_\text{W,max} &= 1948.5 \alpha \epsilon^{-1/2}
\end{align}
For the typical values of $\alpha$ and $\epsilon$ used in this paper the ratio of
\begin{equation}
  \frac{\gamma_\text{E,max}}{\gamma_\text{W,max}} = 0.1 \alpha^{-1/3} \epsilon^{1/6}
\end{equation}
is larger than unity as in the original paper. Also the ordering of modulational instability and non-linear Landau damping are conserved.

\section{Results}

\subsection{Basic simulation}

The simulation setup was first tested with a basic simulation: $\epsilon = 1$, $\alpha= 2.5\cdot 10^{-5}$, $\gamma=10$.
This setup can be regarded as an extreme scenario, since in typical physical environments the beam strength is expected to be smaller than the thermal energy of the background.

\begin{figure}
	\includegraphics[width=0.9\columnwidth]{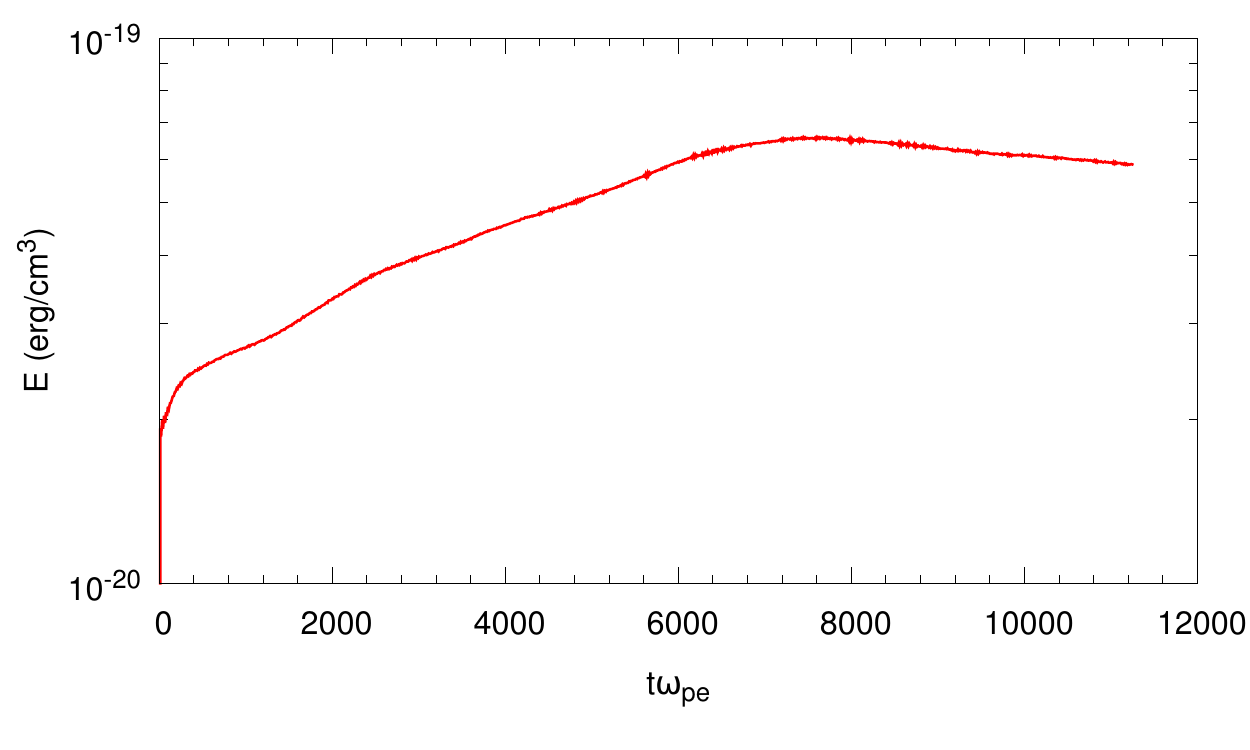}
	\caption{Electric field energy $\epsilon = 1$, $\alpha= 2.5\cdot 10^{-5}$, $\gamma=10$}
	\label{fig:e1g10-energy}
\end{figure}

We have identified several observables as relevant for the development of the instability and the evolution of the distribution function.
In Fig. \ref{fig:e1g10-energy} the energy of the electric field is plotted over time.
The magnetic field energy is following basically the same curve. A sharp increase can be seen until $t=200 \ipt$ which then relaxes until the electric field energy saturates at around $t=7000 \ipt$.\\ The growth rates are about $\gamma=3\cdot 10^{-4}\epf$ at the beginning and growing to $\gamma=7\cdot 10^{-4}\epf$. The analytical values predict a maximum growth rate of $\gamma_E=1.9\cdot 10^{-2}\epf$ for the electrostatic instability and $\gamma_W=3.8\cdot 10^{-3}\epf$ for the aperiodic fluctuation, while the modulation instability should kick in at $t=900 \ipt$.

\begin{figure}
	\includegraphics[width=0.9\columnwidth]{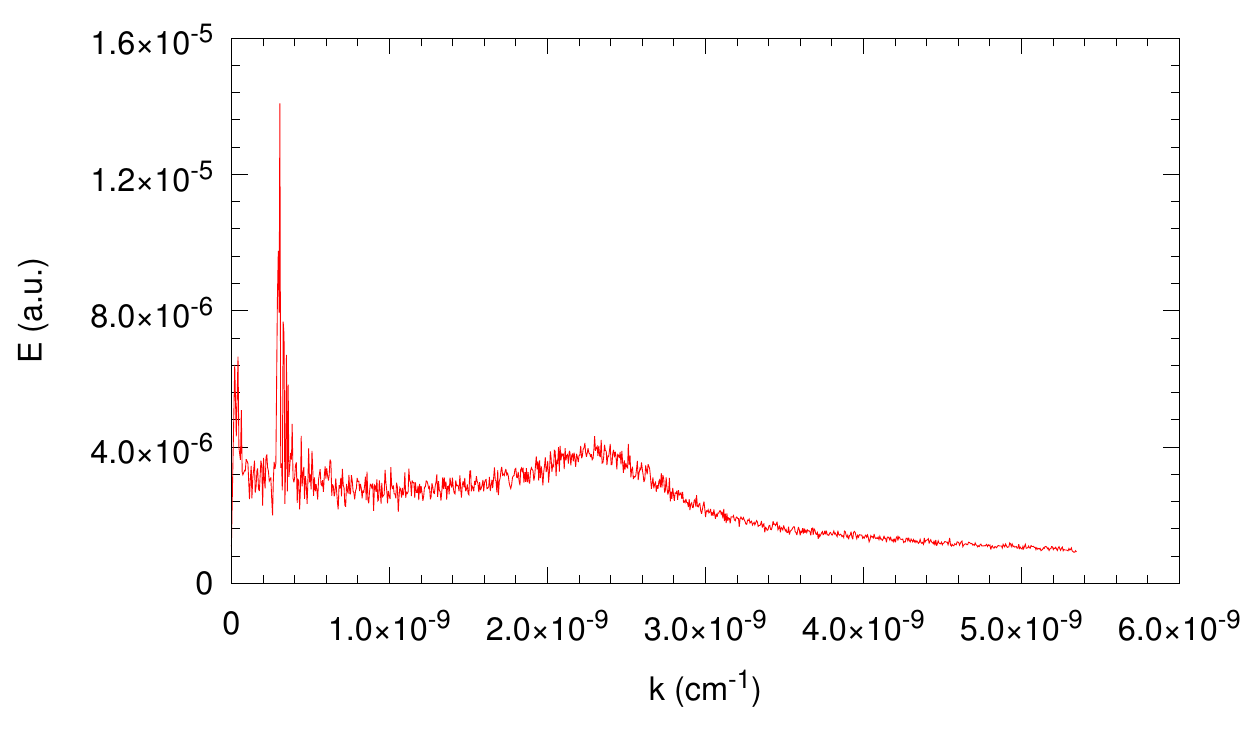}
	\caption{E-field 1D Fourier $\epsilon = 1$, $\alpha= 2.5\cdot 10^{-5}$, $\gamma=10$ at $t=8435\ipt$}
	\label{fig:e1g10-E1d}
\end{figure}

\begin{figure}
	\includegraphics[width=0.9\columnwidth]{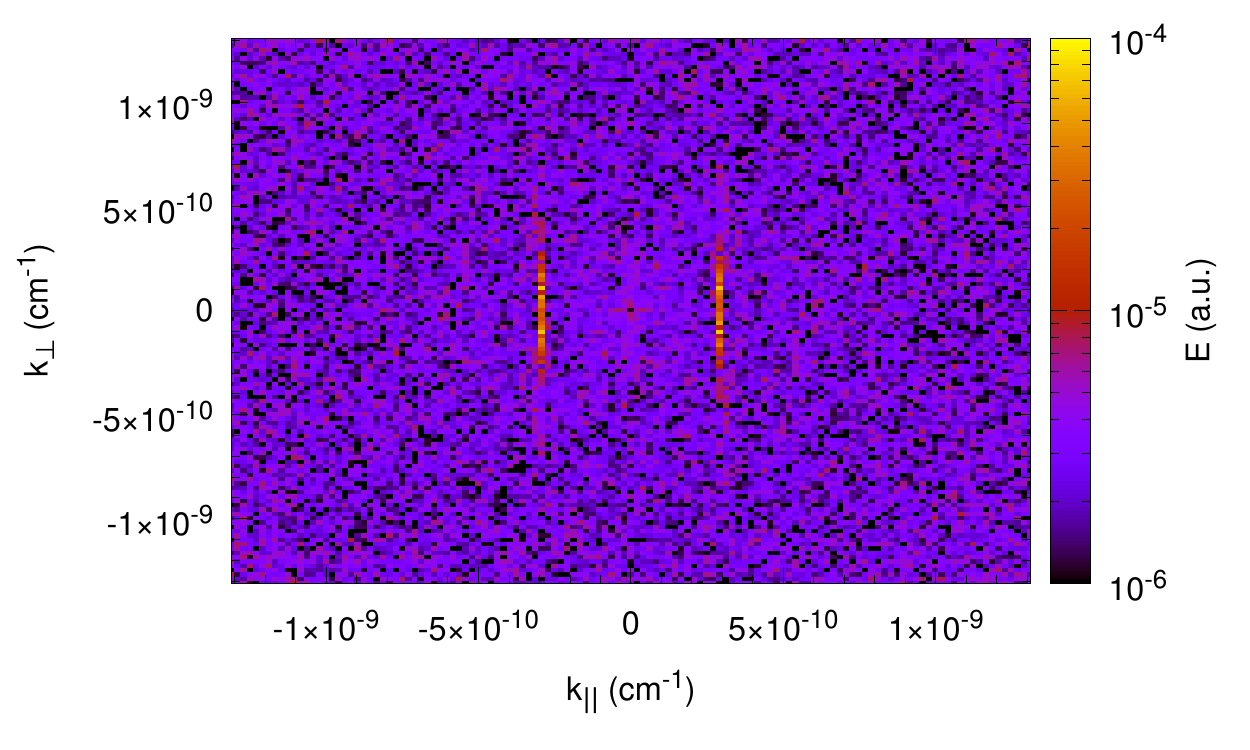}
	\caption{E-field 2D Fourier $\epsilon = 1$, $\alpha= 2.5\cdot 10^{-5}$, $\gamma=10$ at $t=8435\ipt$}
	\label{fig:e1g10-E}
\end{figure}

It should be noted on the one hand, that the growth rates are really only maximum growth rates, so that in this respect the observed growth rates are in line with the analytical predictions, on the other hand the average growth rates in the simulations are in a similar range as the analytical calculations. A detailed look at Fig. \ref{fig:e1g10-E1d} highlights a sharp peak of the Fourier transformed electric field energy at $4\cdot 10^{-10}$ cm$^{-1}$ and another bump at $2.3\cdot 10^{-10}$ cm$^{-1}$. The 2D Fourier transform shows (Fig. \ref{fig:e1g10-E}), that the former peak is linked to a structure at fixed $k_\parallel$ over a wide range of $k_\perp$. This is not predicted by \citet{2012ApJ...752...23C}, who claim that the maximum growth occurs at a fixed angle independent of the absolute of $k$. The position in $k$-space is compatible to $k_\parallel u \approx k_\parallel c = \omega_{pe}$.

\begin{figure}
	\includegraphics[width=0.9\columnwidth]{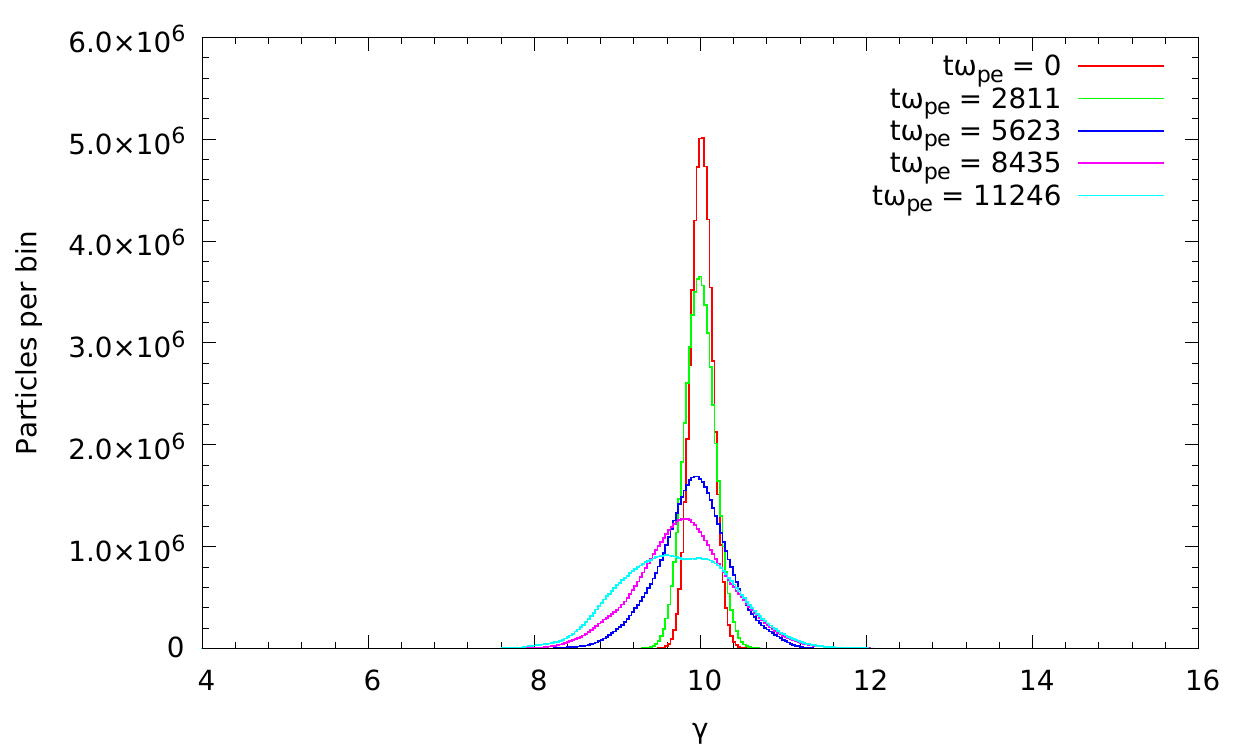}
	\caption{Particle histogram $\epsilon = 1$, $\alpha= 2.5\cdot 10^{-5}$, $\gamma=10$}
	\label{fig:e1g10-histo}
\end{figure}

When observing the change of the distribution function as outlined in Fig. \ref{fig:e1g10-histo}, it can be seen that the peak of the beam is decreasing in amplitude, but even after more than $10^4$ plasma timescales it just moves to a plateau. The prediction of \citet{2012ApJ...752...23C}, which invokes results by \citet{1975AuJPh..28..731G}, says that after around 5000 plasma timescales the peak should have vanished. We conclude that the plateauing of the distribution function slows down the instability, which brings the system to a more or less stable situation after around 7000 plasma timescales.

\subsection{Faster beams}

\begin{figure}
	  \includegraphics[width=0.9\columnwidth]{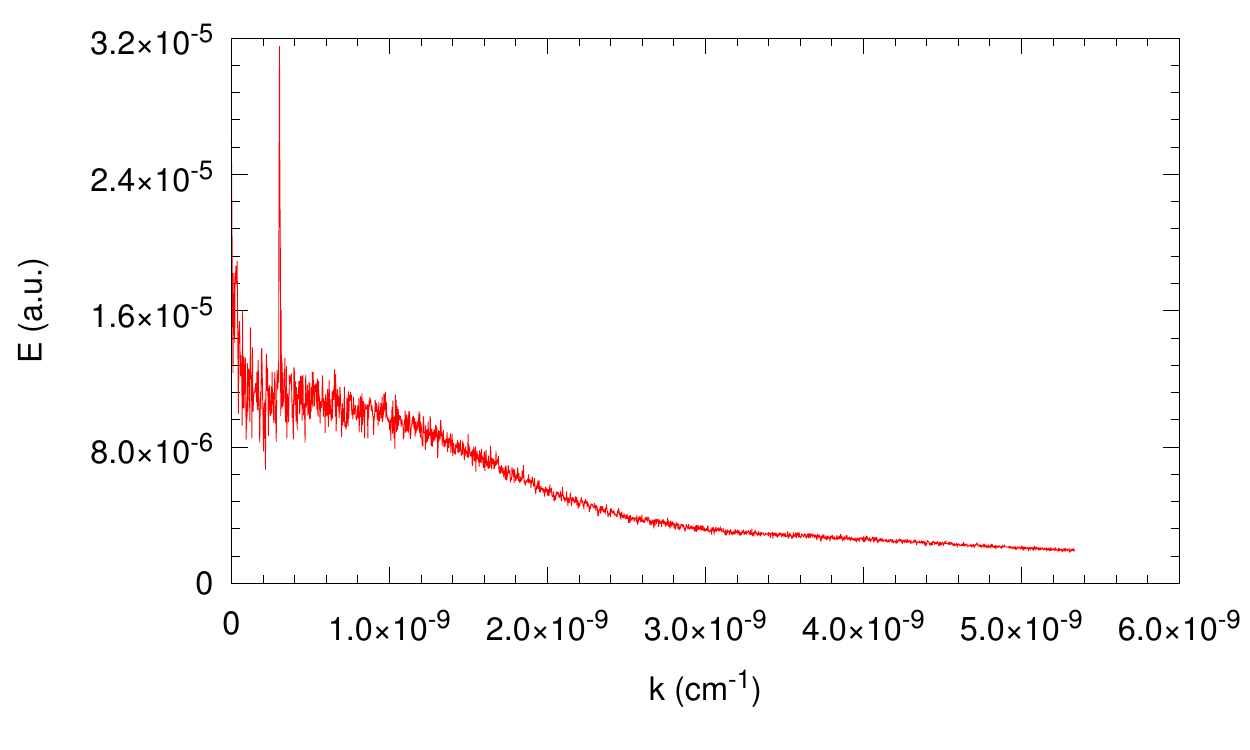}
	  \caption{Faster beam ($\epsilon = 1$, $\alpha= 1.25\cdot 10^{-5}$, $\gamma=20$), E-field 1D Fourier, resolution $1024\times 1024$ at $t=8435\ipt$}
	  \label{fig:e1g20-E1d}
\end{figure}

\begin{figure}
	  \includegraphics[width=0.9\columnwidth]{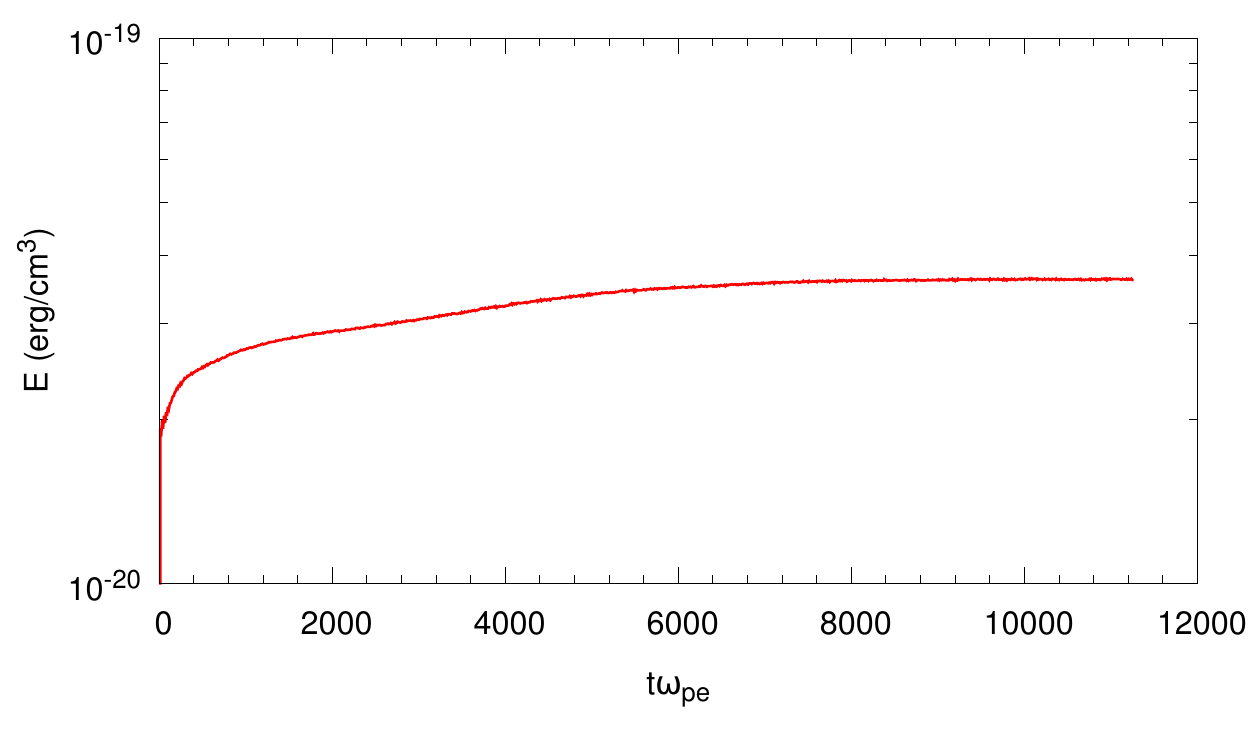}
	  \caption{Faster beam ($\epsilon = 1$, $\alpha= 1.25\cdot 10^{-5}$, $\gamma=20$), Electric field energy, resolution $1024\times 1024$}
	  \label{fig:e1g20-energy}
\end{figure}

\begin{figure}
	  \includegraphics[width=0.9\columnwidth]{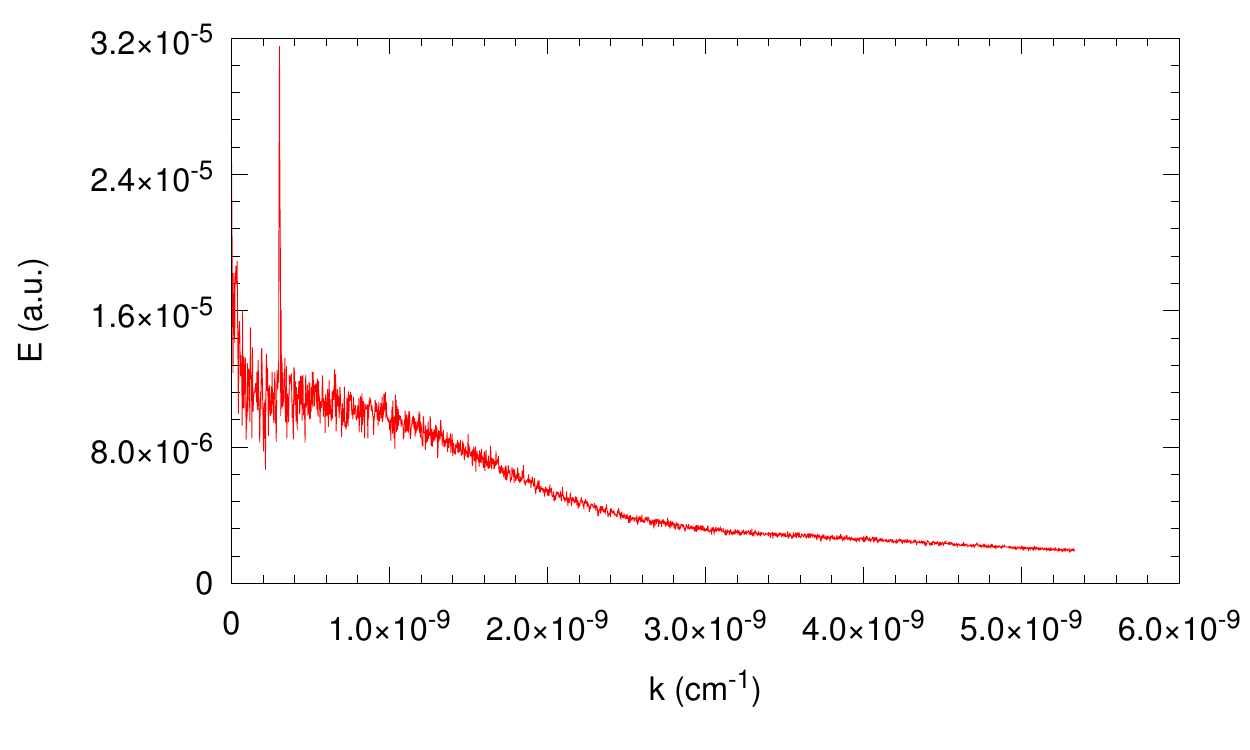}
	  \caption{Faster beam ($\epsilon = 1$, $\alpha= 1.25\cdot 10^{-5}$, $\gamma=20$), E-field 1D Fourier, resolution $2048\times 2048$  at $t=8435\ipt$}
	  \label{fig:e1g20-E1d-big}
\end{figure}

\begin{figure}
	  \includegraphics[width=0.9\columnwidth]{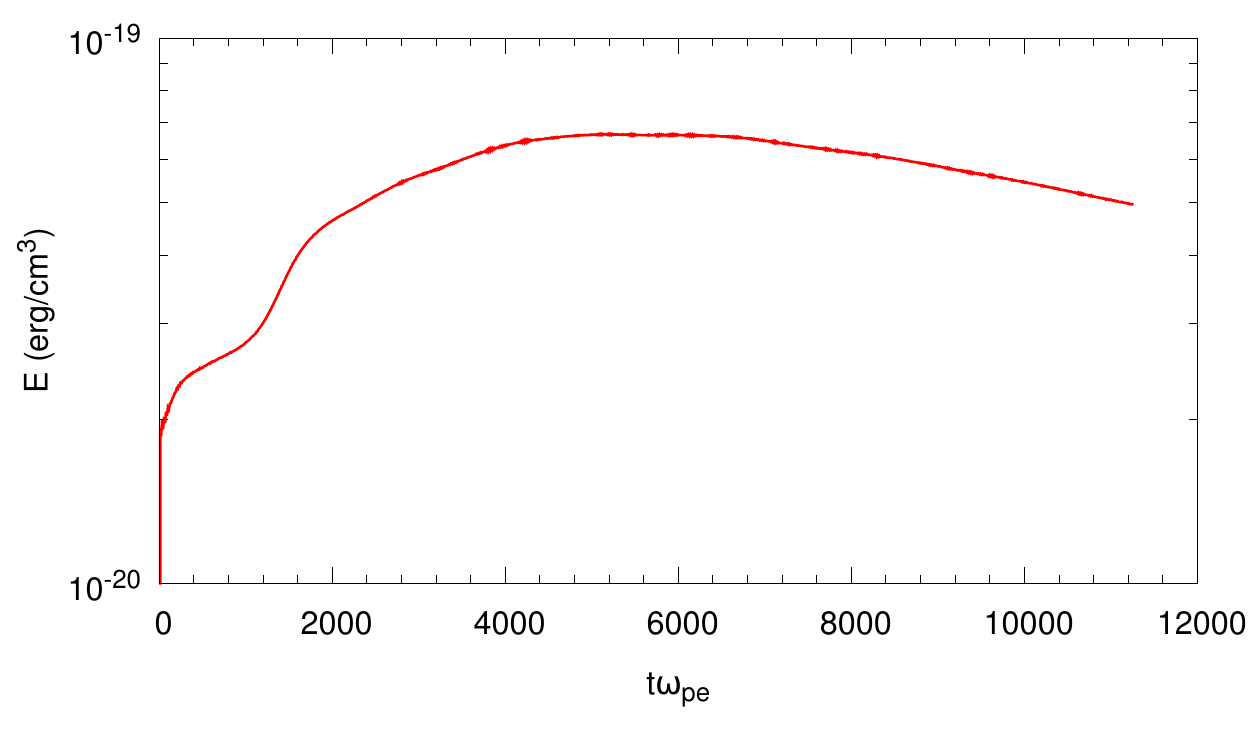}
	  \caption{Faster beam ($\epsilon = 1$, $\alpha= 1.25\cdot 10^{-5}$, $\gamma=20$), Electric field energy, resolution $2048\times 2048$}
	  \label{fig:e1g20-energy-big}
\end{figure}

\begin{figure}
	  \includegraphics[width=0.9\columnwidth]{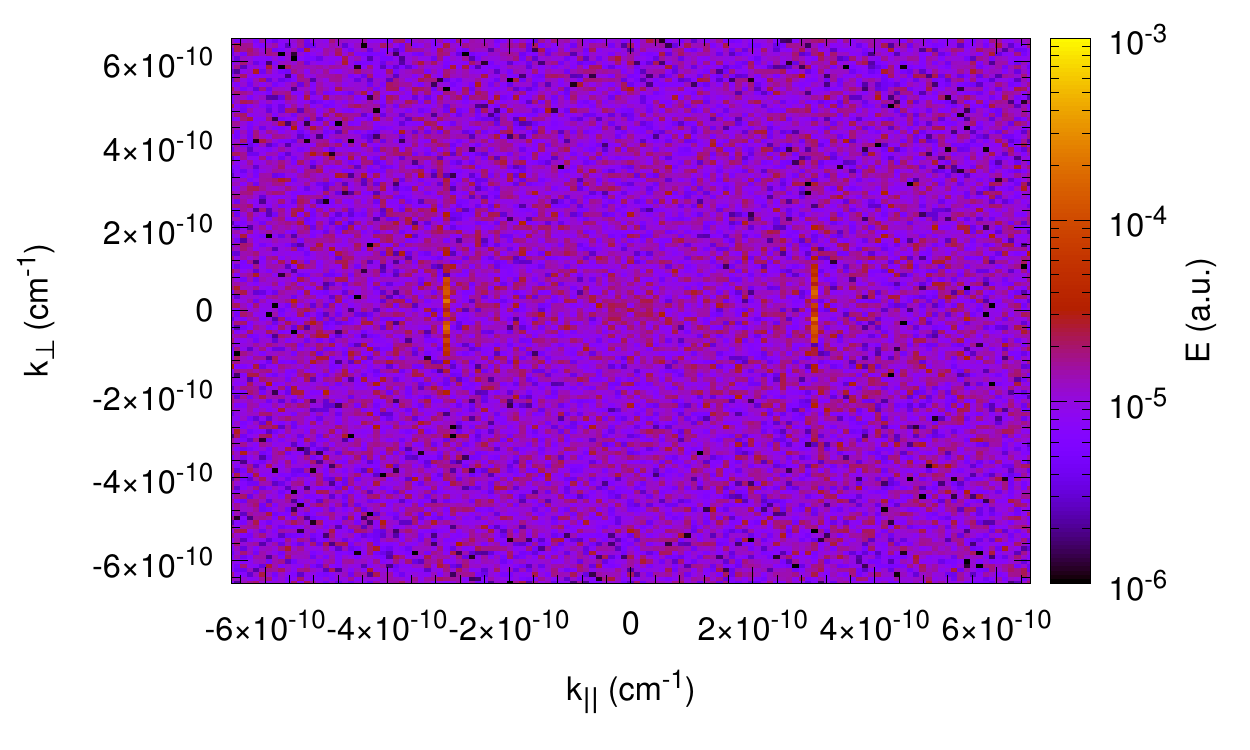}
	  \caption{Faster beam, E-field 2D Fourier, resolution $2048\times 2048$  at $t=8435\ipt$}
	  \label{fig:e1g20-E-big}
\end{figure}

Since the basic simulation is limited to a beam Lorentz factor of only 10, simulations with faster beams are necessary to make statements about the evolution of beams in the actual physical setting.
Our setting for a mildly faster beam was $\epsilon = 1$, $\alpha= 1.25\cdot 10^{-5}$, $\gamma=20$.
First attempts with a similar resolution (Figs. \ref{fig:e1g20-E1d}, \ref{fig:e1g20-energy}) led to results showing a significantly different behavior compared to the slower beam.
It turned out that faster beams require a higher resolution.
The results analogously to the $\gamma=10$ case are shown in Figs. \ref{fig:e1g20-E1d-big}, \ref{fig:e1g20-energy-big}, \ref{fig:e1g20-E-big}.

When calculating the theoretically predicted values, the results are only mildly different to the basic case:  $\gamma_E=1.2\cdot 10^{-2}\epf$ for the electrostatic instability and $\gamma_W=1.9\cdot 10^{-3}\epf$ for the aperiodic fluctuation. For the well resolved simulation there is a decline of the electric field energy at around 6000 plasma timescales, but the distribution function changes basically in the same way as before.


\subsection{Strong beam}

From the scenario of the faster beam we developed a strong beam scenario which also has a beam with Lorentz factor $\gamma=20$, but a different energy density ratio.
We assumed $\epsilon=10$ with a constant density ratio of  $\alpha= 1.25\cdot 10^{-4}$ and again with a resolution of $2048\times2048$.

\begin{figure}
	  \includegraphics[width=0.9\columnwidth]{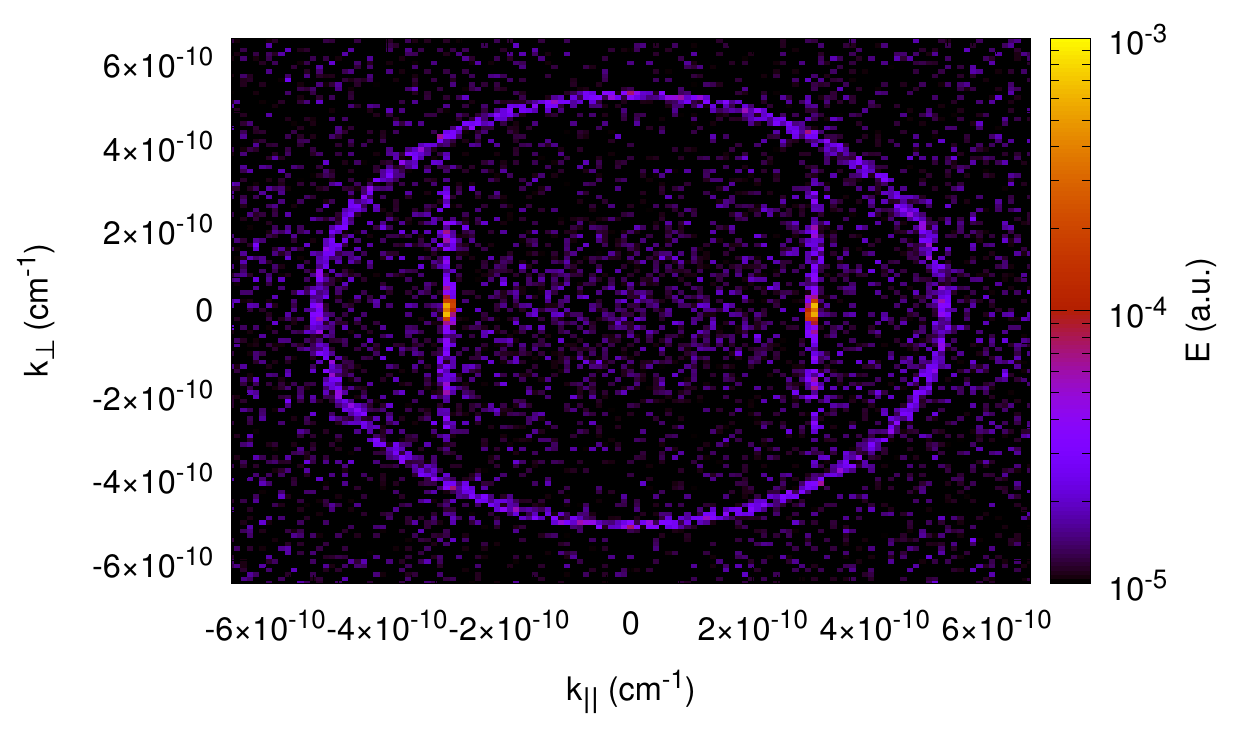}
	  \caption{Strong beam ($\epsilon = 10$, $\alpha= 1.25\cdot 10^{-4}$, $\gamma=20$ at $t=8435\ipt$), E-field 2D Fourier}
	  \label{fig:e10g20-E-big}
\end{figure}

The results for this simulation run differ strongly from the previous simulation: The growth rates are much higher than in the basic simulation ($7\cdot 10^{-4} \omega_{pe}$ rising to $7\cdot 10^{-3} \omega_{pe}$). This is only partially expected from the theoretical calculations. The expected growth rate for the electrostatic fluctuations are only marginally higher ($2.6\cdot 10^{-2}\omega_{pe}$), while the aperiodic fluctuation rate is even decreased ($6\cdot 10^{-4}\omega_{pe}$). A close look at the 2D Fourier transform in Fig. \ref{fig:e10g20-E-big} shows the already well known features at $k_\parallel = \SI{3e-10}{\per\centi\meter}$, but additionally it shows a ringlike structure at $|k| = \SI{5e-10}{\per\centi\meter}$. The fact that this feature only appears for $\epsilon > 1$, which can be considered as unphysical scenario, puts the results of \citet{Sironi_2014} into question.

\begin{figure}
	  \includegraphics[width=0.9\columnwidth]{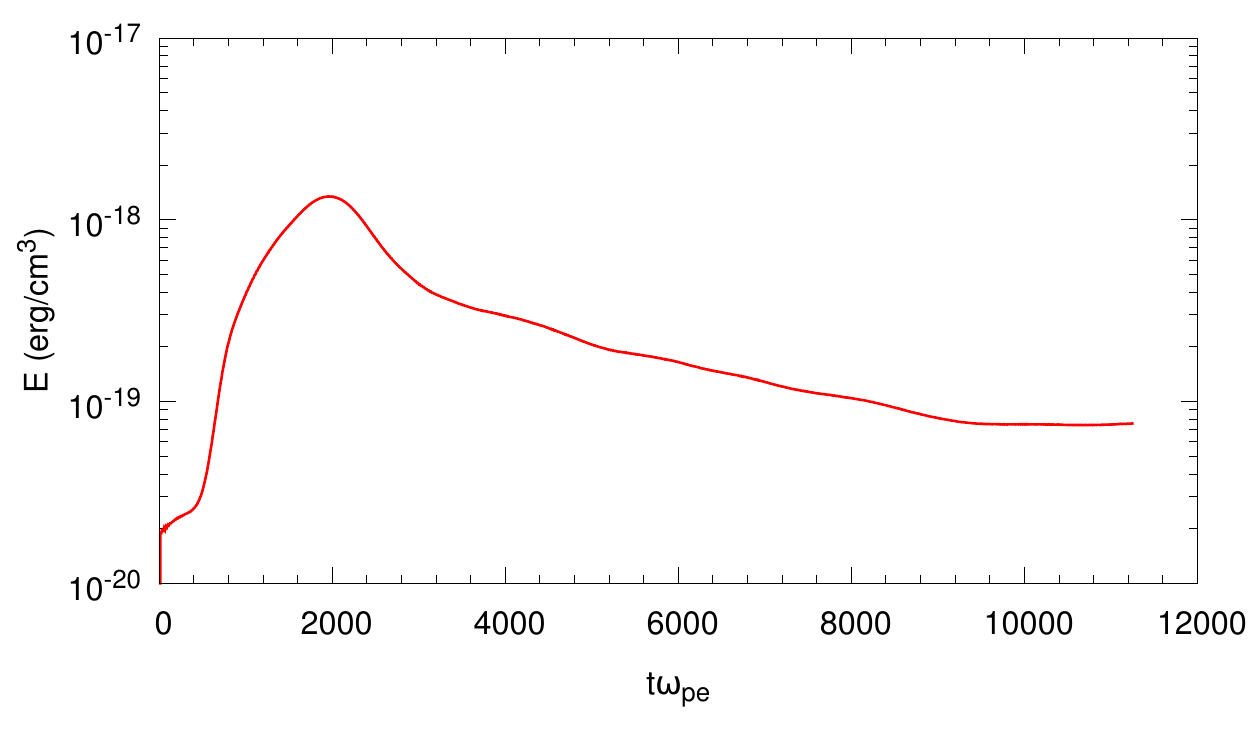}
	  \caption{Strong beam ($\epsilon = 10$, $\alpha= 1.25\cdot 10^{-4}$, $\gamma=20$), Electric field energy}
	  \label{fig:e10g20-energy-big}
\end{figure}

\begin{figure}
	  \includegraphics[width=0.9\columnwidth]{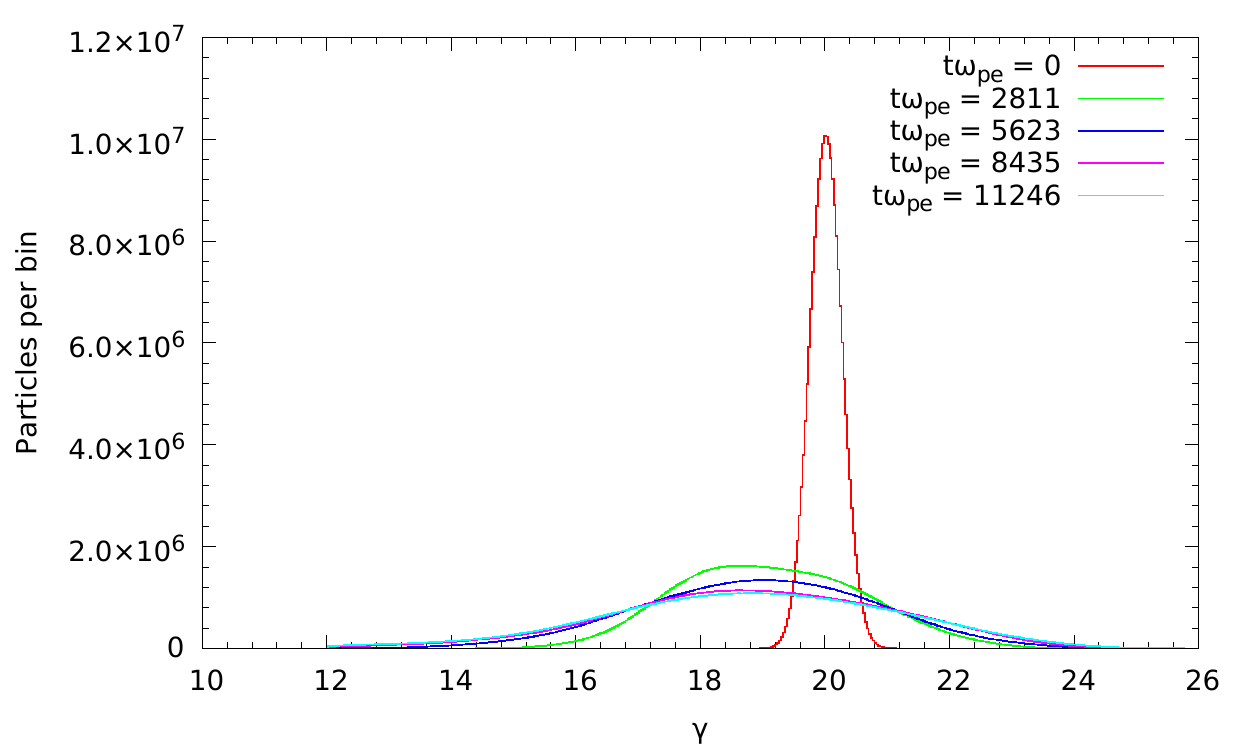}
	  \caption{Strong beam ($\epsilon = 10$, $\alpha= 1.25\cdot 10^{-4}$, $\gamma=20$), Particle histogram}
	  \label{fig:e10g20-histo-big}
\end{figure}

An even more interesting result can be seen, when inspecting the histogram in Fig. \ref{fig:e10g20-histo-big}. The peak shows a sharp decrease in amplitude (which would only be a qualitative difference to the basic simulation), but it also shows a shift in the peak position. This is a quantitative difference and the only case in this series of simulations, where the beam suffers an actual loss of energy.

\subsection{Weak beam}

The last setup is the weak beam with $\epsilon = 0.1$, $\alpha= 2.5\cdot 10^{-6}$, $\gamma=10$.
This is a mostly realistic scenario with regard to the energy density ratio.

\begin{figure}
	  \includegraphics[width=0.9\columnwidth]{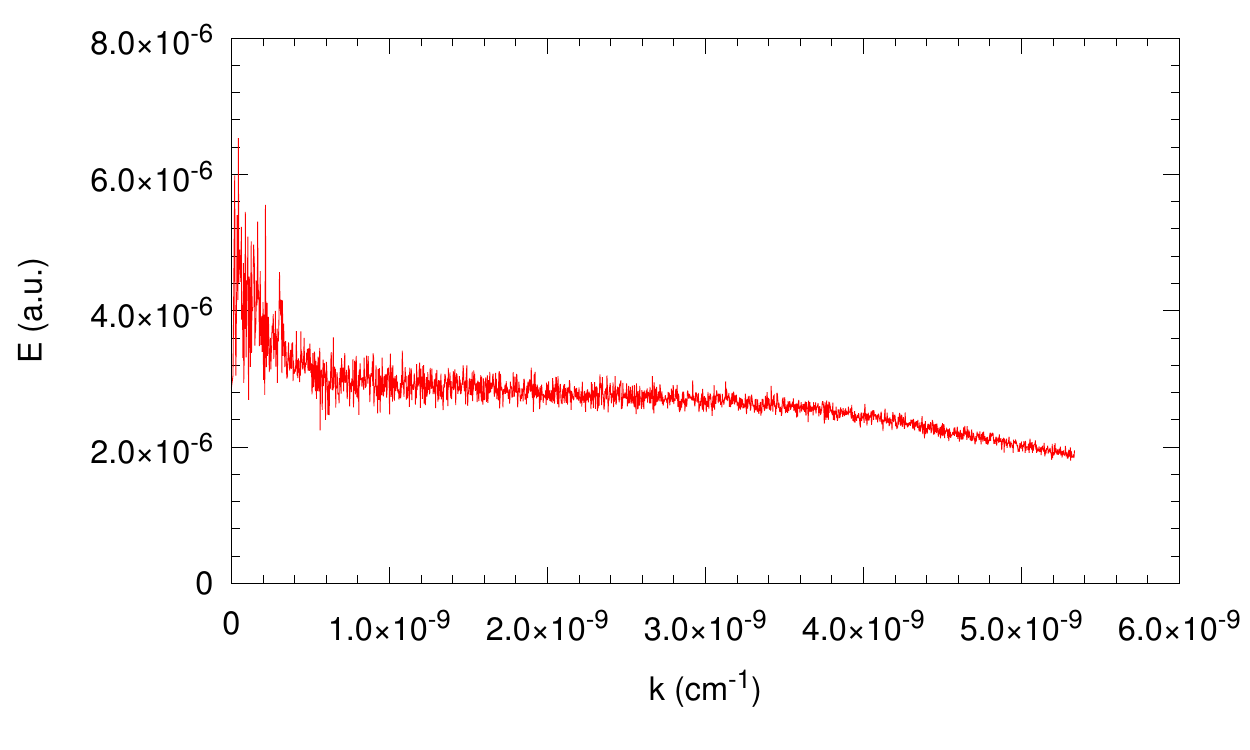}
	  \caption{Weak beam  $\epsilon = 0.1$, $\alpha= 2.5\cdot 10^{-6}$, $\gamma=10$, E-field 1D Fourier at $t=8435\ipt$}
	  \label{fig:e01g10-E1d-big}
\end{figure}

\begin{figure}
	  \includegraphics[width=0.9\columnwidth]{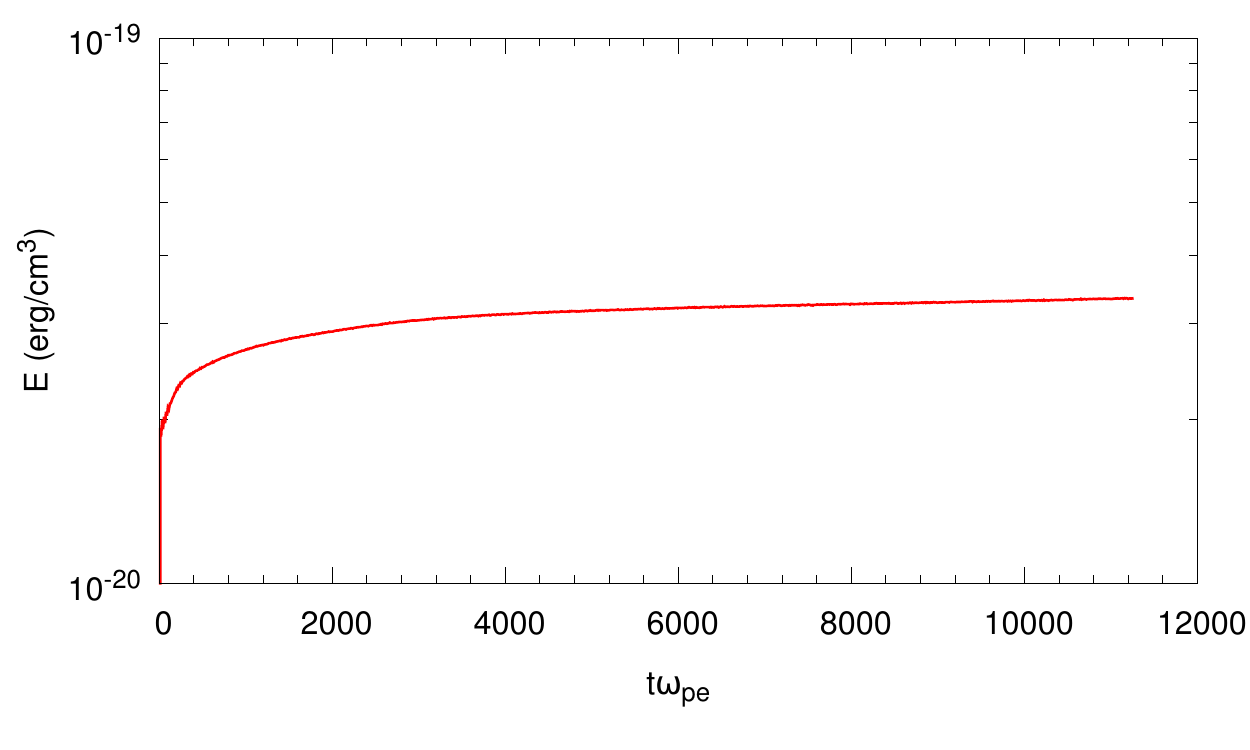}
	  \caption{Weak beam $\epsilon = 0.1$, $\alpha= 2.5\cdot 10^{-6}$, $\gamma=10$, Electric field energy}
	  \label{fig:e01g10-energy-big}
\end{figure}

\begin{figure}
	  \includegraphics[width=0.9\columnwidth]{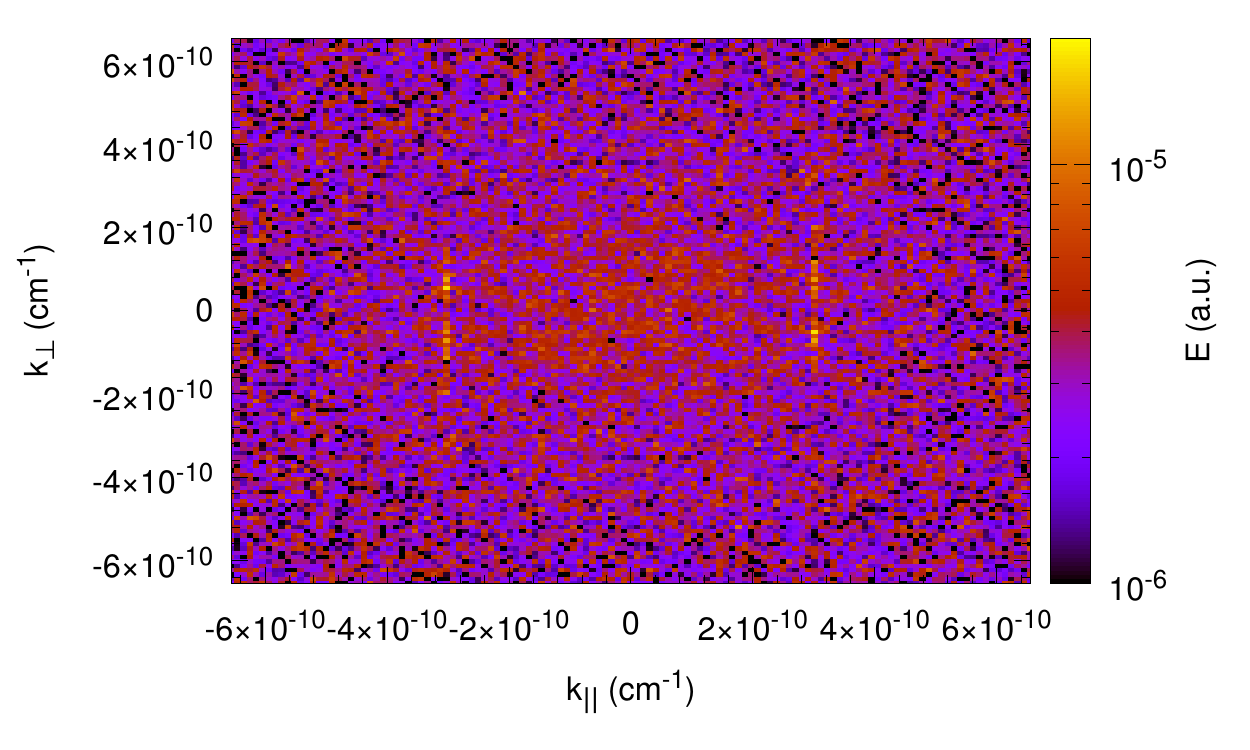}
	  \caption{Weak beam $\epsilon = 0.1$, $\alpha= 2.5\cdot 10^{-6}$, $\gamma=10$, E-field 2D Fourier at $t=8435\ipt$}
	  \label{fig:e01g10-E-big}
\end{figure}

The observed growth rate (Fig. \ref{fig:e01g10-energy-big}) of the instability here is approximately $2\cdot 10^{-4} \epf$ with theoretical values of $8.8\cdot 10^{-3} \epf$ (electrostatic) and $1.2\cdot 10^{-3}\epf$ (aperiodic). A saturation stage is reached very early. The 1D Fourier transform (Fig. \ref{fig:e01g10-E1d-big}) and 2D Fourier transform (Fig. \ref{fig:e01g10-E-big}) are similar to the basic simulation, but the instability peak is much less pronounced.

\begin{figure}
	  \includegraphics[width=0.9\columnwidth]{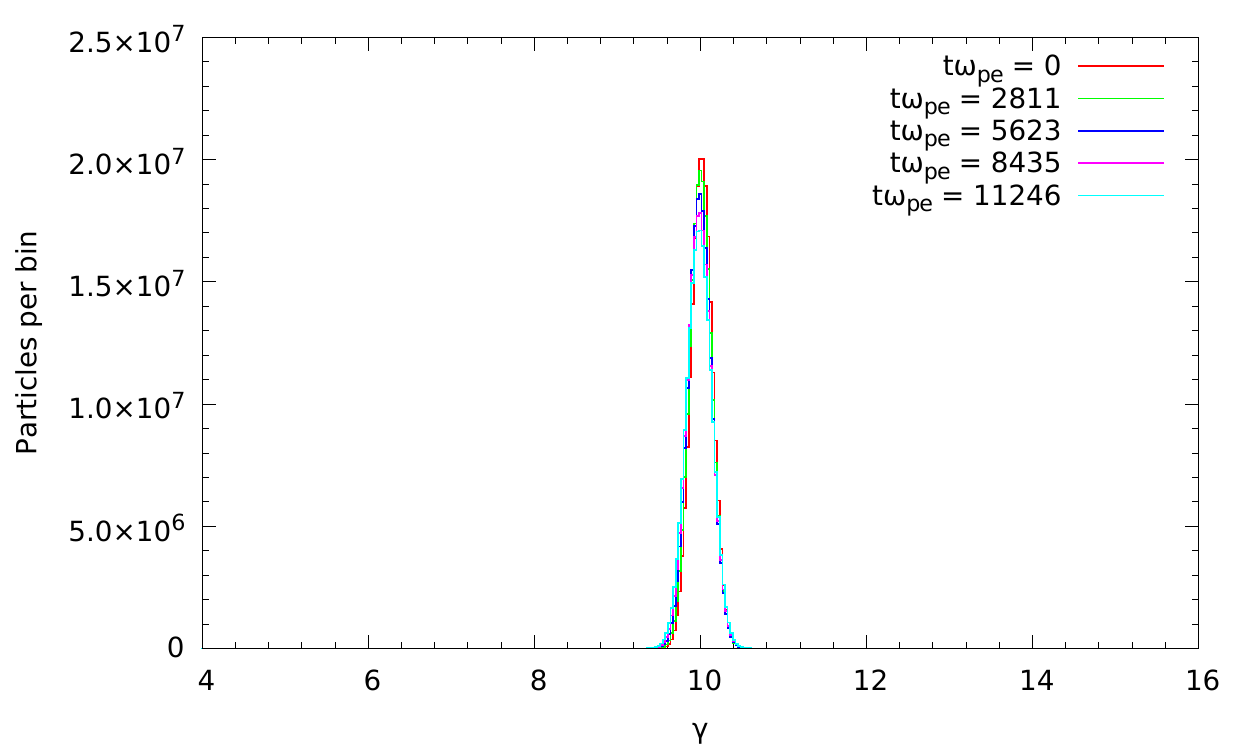}
	  \caption{Weak beam $\epsilon = 0.1$, $\alpha= 2.5\cdot 10^{-6}$, $\gamma=10$, Particle histogram}
	  \label{fig:e01g10-histo-big}
\end{figure}

An important feature seen here is the negligible change of the pair beam distribution function as shown in Fig. \ref{fig:e01g10-histo-big}.

\section{Discussion}

In the simulations presented in this article we could recover the onset of an instability caused by electron-positron pair beams.
The general existence of this instability is not affected by the change of the energy density ratio of beam and background.
The actual development of the instability depends strongly on the energy density ratio $\epsilon$.

Common to all simulations, regardless of the beam Lorentz factor $\gamma$ and the energy density ratio $\epsilon$ is the excitation  at $k_\parallel = \SI{3e-10}{\per\centi\meter}$ over a whole range of perpendicular wave lengths $k_\perp$. The strength of this emission depends on the beam strength (stronger beams lead to stronger emission) and the beam speed (faster beams produce stronger emission).
The two opposite cases of strong and weak beams produce interesting opposing results:
Due to the weakness of the beam Fig. \ref{fig:e01g10-E-big} is mostly dominated by noise, while the strong beam produces a clear signal along the described $k_\parallel$ axis as seen in Fig. \ref{fig:e10g20-E-big}.
An additional feature seen there is a ring like structure in the $k$-space with a radius of $|k| = \SI{5e-10}{\per\centi\meter}$.

The one-dimensional energy spectra reflect this finding: Regardless of $\epsilon$ and $\gamma$ they all show falling spectra with a peak around $|k|=3\cdot 10^{-10}$ cm$^{-1}$. In the case of $\epsilon=10$ another peak related to the ring-like structure can be seen. The underresolved case in Fig. \ref{fig:e1g20-E1d} shows an additional hump at $|k|=2.2\cdot 10^{-10}$ cm$^{-1}$, that vanishes when increasing the velocity. This simulation has, therefore, been rejected as unphysical.

One important fact to note is, that in all simulations a linear stage can be observed  on timescales shorter than $100 \omega_{pe}^{-1}$ after that the rise of the instability slows down, eventually reaching a maximum.
For the case of strong and fast beam there is even a decline of electric field energy observable.
The difference between these different runs may be explained by taking a look at the distribution functions.

For the basic simulations we see a broadening of the peak distribution function after a rather long time scale of more than $5000 \omega_{pe}^{-1}$.
This spread continues, but even after $11000 \omega_{pe}^{-1}$ the peak has only broadened by \SI{20}{\percent} and the peak amplitude is still a fifth of the initial amplitude.
When this is compared to the strong beam scenario, it becomes clear that the evolution of the distribution function happens much faster and changes the distribution function much more drastically: Not only has the distribution function changed after less than $3000 \omega_{pe}^{-1}$ but also the position of the maximum has shifted.
This may be linked to the strong decline of the instability observed in the electric field.

The weak beam scenario shows a drastically different behavior: Even after more than $8000 \omega_{pe}^{-1}$ the distribution function of beam pairs is almost completely unchanged.
We want to stress the fact that we resolve the timescales for the modulation instability and non-linear Landau damping in this simulation \citep[cf.][]{2012ApJ...758..102S}.

We conclude that for the simulation setup presented here, the change of the distribution is negligible over longer times, when the energy density ratio is below 1.
For actual physical scenarios we would expect typically even lower values of $\epsilon$ than those shown here.
The loss channels of inverse Compton scattering and magnetic field deflection are, therefore, still open for the pair beams. This is consistent with assumptions that are usually made in the determination of the intergalactic magnetic field based on observations of distant point sources.

Due to the high numerical effort involved, our simulations are limited to low $\gamma$ values, but as the comparison of the basic scenario and the fast beam show, $\epsilon$ is the far more decisive parameter in the evolution of the system as such.

\section{Conclusion}

We have shown simulations of systems containing a hot, thermal proton-electron plasma and mildly relativistic electron-positron beams. Our simulations suggest that for low energy density ratios (i.e., less energetic beams) instabilities are created, which do not lead to a strong change in the distribution function of the beam.\\
When taking into account the physical scenario, we would conclude, that while the instabilities in question may broaden the beam distribution, they do not provide enough energy loss to explain missing GeV photons. This brings us back to the original paper by \citet{2010Sci...328...73N}: We do not observe upscattered GeV photons from EBL generated pair beams and the instability does not successfully remove the beam electrons, therefore, magnetic deflection may still be the governing process.

\section*{Acknowledgments}
We acknowledge the use of the \emph{ACRONYM} code and would like to thank the developers (Verein zur F\"orderung kinetischer Plasmasimulationen e.V.) for their support.

The work of AK was supported by the Deutsche Forschungsgemeinschaft through grant Schl 201/31-1.

FS acknowledges support from NRF through the MWL program. This work is based upon research supported by the National Research Foundation and Department of Science and Technology. Any opinion, findings and conclusions or recommendations expressed in this material are those of the authors and therefore the NRF and DST do not accept any liability in regard thereto.

\bibliographystyle{aa}
\bibliography{apj-jour,dinge}

\end{document}